\documentstyle[12pt,epsfig]{article}  
\begin{document}

%===================> ADD here your LATEX definitions

%========================================================================% 

%################################################## titlepage declaration

\begin{titlepage}

\pagenumbering{arabic}
\vspace*{-1.5cm}
\begin{tabular*}{15.cm}{l@{\extracolsep{\fill}}r}
{\bf DELPHI Collaboration} & 
%===================> DELPHI note number       =====> To be filled <=====%
DELPHI 98-42 PHYS 772 
%========================================================================%
\\
& 
%===================> DELPHI note date         =====> To be filled <=====%
27 April, 1988
%========================================================================%
\\
&\\ \hline
\end{tabular*}
\vspace*{2.cm}
\begin{center}
\Large 
{\bf
%===================> DELPHI note title        =====> To be filled <=====%
 Study of $b \rightarrow u \ell \bar \nu$ Decays \\
with an Inclusive Generator 
%========================================================================%
} \\
\vspace*{2.cm}
\normalsize { 
%===================> DELPHI note author list  =====> To be filled <=====%
   {\bf M.~Battaglia}\\
   {\footnotesize Dept. of Physics, \\
      University of Helsinki (Finland)}\\
   
%========================================================================%
}
%\vspace*{2.cm}
\end{center}
\vspace{\fill}
\begin{abstract}
\noindent
%===================> DELPHI note abstract     =====> To be filled <=====%
A generator for inclusive $b \rightarrow u \ell \bar \nu$ decays has been 
developed. 
Different prescriptions have been applied to describe the kinematics of the 
$b$ quark inside the hadron and the generation of the hadronic final states.
Results are presented with particular attention to the invariant mass of the
hadronic system recoiling against the lepton and its resonance decomposition. 
These studies are of special relevance for the extraction of $|V_{ub}|$ from 
semileptonic $B$ decays at LEP and at $B$ factories. 
%=========================================================================%

\end{abstract}
\vspace{\fill}
%\begin{center}
%==========> Proceedings.. presented at ..==> To be filled if needed<=====%

%=========================================================================%
%\end{center}
\vspace{\fill}
\end{titlepage}

%\pagebreak

%\begin{titlepage}
%\mbox{}
%\end{titlepage}
\newpage

\mbox{  }

%\pagebreak
\newpage

\setcounter{page}{1}    

%##################################################################### Text

%==================> DELPHI note text          =====> To be filled <======%
\section{Introduction}

The measurement of the branching ratio for the decay 
$b \rightarrow u \ell \bar \nu$ provides the cleanest way to determine the 
$|V_{ub}|$ element in the CKM mixing matrix. Evidence of the non-zero
value of $|V_{ub}|$ has been first obtained by both ARGUS and CLEO by 
observing leptons produced in $B$ decays with momentum exceeding the 
kinematical limit for $b \rightarrow c \ell \bar \nu$ transitions~\cite{cleo1,
argus1}. 
The extraction of $|V_{ub}|$ from the yield of leptons above the 
$b \rightarrow c \ell \bar \nu$ endpoint is subject to large systematical 
uncertainties. More recently, exclusive $B \rightarrow \pi \ell \bar \nu$ and 
$B \rightarrow \rho \ell \bar \nu$ decays have been observed by CLEO and their 
rates measured~\cite{cleo2,cleo3}. Still the derivation 
of $|V_{ub}|$ from exclusive semileptonic decays, contrary to the case for
$|V_{cb}|$ in $B \rightarrow D^* \ell \bar \nu$, has significant model 
dependence.

The extraction of $|V_{ub}|$ from the shape of the invariant mass of the 
hadronic system recoiling against the lepton in $b \rightarrow u \ell \bar \nu$
transitions was proposed several years ago~\cite{barger} and it has been 
recently the subject of new studies~\cite{falk,bdu}. 
The proposed method starts from the observation that the hadronic system 
recoiling against the lepton in the decay has invariant mass lower than the 
charm mass for the majority of $b \rightarrow u \ell \bar \nu$ decays. The 
model dependence in predicting the shape of this invariant mass distribution 
has been claimed to be under control within about $10-15~\%$ if 
$b \rightarrow u \ell \bar \nu$ decays can be distinguished from the 
$b \rightarrow c \ell \bar \nu$ ones for masses of the hadronic system up to
cut values close enough to the $D$ mass~\cite{bdu}. This corresponds to a
model uncertainty of 5-7\% on the extraction of $|V_{ub}|$.

The predicted shape of the invariant mass distribution depends mainly on the 
kinematics 
of the heavy and spectator quarks inside the $B$ hadron and on the quark 
masses. However from the experimental point of view, the hadronisation process,
transforming the $u \bar q$ system into the observable hadronic final state, 
represents a significant source of additional model uncertainties. 

Several models have been proposed to describe both steps of the decay process.
This note summarizes the results of a study performed by developing a 
dedicated $b \rightarrow u \ell \bar \nu$ decay generator, BTOOL. 
This generator implements different prescriptions for the initial state 
kinematics and the resonance decomposition of the hadronic final states. Its 
results are used to define the model dependent systematics in the extraction 
of $|V_{ub}|$ from the hadronic mass spectrum in semileptonic $B$ decays.

\section{The BTOOL Generator}

The decay generator provides the four momenta of the stable decay products
in $b \rightarrow u \ell \bar \nu$ transitions. This requires a model for the 
kinematics of the $b$ and spectator $\bar q$ quarks inside the $B$ hadron, the 
description of the $Q^2$ distribution of the virtual $W$ and of the kinematics
in the $b \rightarrow u W$ and $W \rightarrow \ell \bar \nu$ decays, and 
finally a model for the hadronisation of the $u \bar q$ system. 
In the following the 
implementation of the decay in the generator is discussed. In obtain to
study model dependences and systematic effects, different prescriptions have 
been adopted.

\subsection{ACCMM Model}

In the ACCMM model~\cite{accmm} the $B$ hadron consists of the $b$ quark and 
the spectator $\bar q$ quark moving back-to-back in the $B$ rest frame.Their
momenta $p$ are distributed according to the gaussian distribution: 
\begin{eqnarray}
\phi(p) = \frac{4}{\sqrt{\pi}p^3_F} e^{-\frac{p^2}{p^2_F}}
\end{eqnarray}
where the width $p_F$ is known as Fermi motion and represents a
parameter in the model. The normalisation is chosen such that
\begin{eqnarray}
\int^{+\infty}_{0} dp~p^2 \phi(p) = 1.
\end{eqnarray} 
The choice of the $p_F$ parameter and of the mass of the spectator quark
$m_q$ are discussed in details in the next section.

\subsection{Parton Model}

An alternative picture of the $b$ quark kinematics has been proposed as an
application of the parton model to heavy quark decays ~\cite{parton}. 
In this model the decay is considered in a frame where the $B$ hadron moves 
with large momentum (infinite momentum frame or Breit frame). 
In this frame the $b$ 
quark behaves as a free particle carring a fraction $z$ of the $B$ momentum,
$p_b = z p_B$. The functional form of $f(z)$ can be extracted from the 
$b$ fragmentation function since the probability of finding the $b$ quark 
carrying a fraction $z$ of the hadron momentum corresponds to the probability
for a $b$ quark to produce a $B$ hadron with a fraction $z$ of its energy.  
This is usually described by the Peterson fragmentation 
function~\cite{peterson} that has the form:
\begin{eqnarray}
f(z) = \frac{N z(1-z)^2}{((1-z)^2+\epsilon_b z)^2}
\end{eqnarray}
where $\epsilon_b$ is a free parameter. In this way, the parton model offers
an advantage since the kinematics of the $b$ quark is described by a function 
that can be directly related with experimental data on $b$~fragmentation.

\subsection{QCD Universal Structure Function}

Recently there has been progress in defining the Fermi motion in the 
framework of QCD~\cite{neubert1,neubert2,dsu}. 
This has been achieved in terms of a universal structure 
function describing the distribution of the light-cone residual momentum of 
the heavy quark inside the hadron. At leading order and in the large $m_b$
limit, the light-cone residual momentum $k_+$ can be expressed as the 
difference between the b quark pole mass and its effective mass $m^*_b$ inside
the hadron: $m^*_b = m_b + k_+$. As a consequence 
$k_+ < \bar \Lambda = m_B - m_b$. An ansatz for the shape of the universal 
structure function has been suggested~\cite{dsu} in the form:
\begin{eqnarray}
f(z) = z^a (1 - c z) e^{- c z}
\end{eqnarray}
where $z = 1 - \frac{k_+}{\bar \Lambda}$, and the coefficients $a$ and $c$
depend on the values of $\bar \Lambda$ and of the kinetic energy operator
as discussed in the next section.

\subsection{Decay kinematics}

The $b$ quark decays as $b \rightarrow W u$, the $W$ and $u$ being 
emitted back to back in the $b$ rest frame with an isotropic distribution 
of their emission angle w.r.t the $b$ direction. The virtual $W$ is
characterized by an effective mass $Q^2$. The kinematics of
the $b$ and $W$ decays correspond to that for two body decays.
Therefore the model dependence that propagates to the final state kinematics
depends on the choice of the values of $p_F$ and $m_q$ and of the $Q^2$ 
distribution. By defining $x^2 = Q^2/m^2_b$ the differential decay rate can be
expressed as: 
\begin{eqnarray}
\frac{d\Gamma}{dx} = \frac{G^2_F m^5_b |V_{ub}|^2}{192 \pi^3}
(F_0(x) -\frac{2\alpha_s}{3\pi} F_1(x)) 
\end{eqnarray}

The functions $F_0(x)$ and $F_1(x)$ describe the tree-level contribution and
the QCD corrections terms. Following Ref.~\cite{kuhn}, they can be written,
in the limit $m_u \rightarrow 0$, as:
\begin{eqnarray}
F_0(x) = 2(1-x^2)^2(1+2x^2)
\end{eqnarray}
and 
\begin{eqnarray}
F_1(x) = (\pi^2 +2 S_{1,1}(x^2) -2 S_{1,1}(1-x^2)) + 8x^2(1-x^2-2x^4)ln(x) \\
+2(1-x^2)^2(5+4x^2)ln(1-x^2)-(1-x^2)(5+9x^2-6x^4)
\end{eqnarray}
where $S_{1,1}(x)$ is the Nielsen polylogarithm.
The resulting $Q^2$ distribution is shown in Figure~1.
\begin{figure}[h!]
\begin{center}
\epsfig{file=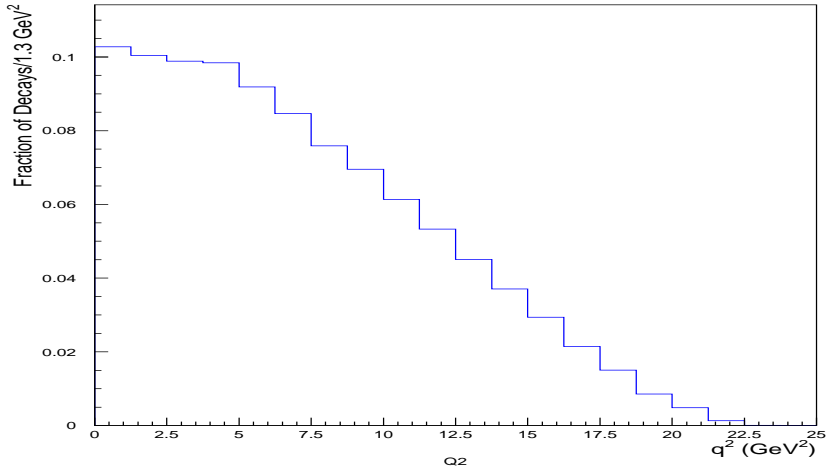,width=12.0cm,height=7.0cm,clip=}
\end{center}
\caption[]{\sl The $q^2$ distribution in the decay 
$b \rightarrow u \ell \bar \nu$ generated according to Eq.~(5).}
\label{fig:q2}
\end{figure}

The virtual $W$ is forced to decay to the lepton-neutrino pair. 
Due to the spin of the $W$, its decay is not isotropic and the angular 
distribution of the lepton varies as $1 + cos^2\theta$.
In computing the lepton and neutrino momenta in the $W$ rest frame the 
lepton mass is taken into account. 
After the decay the lepton and neutrino are boosted back to the $B$
rest frame. The $u$ quark is also boosted to the same system, and
at this point the hadronic system is treated. 

\subsection{The hadronic system}

The energy and invariant mass of the hadronic system correspond to those of 
the $u \bar q$ quark pair. The energy and invariant mass at the parton level 
can be compared with predictions obtained in QCD and heavy quark expansion.
The observable final states are then generated by applying different
prescriptions for describing the evolution of the quark fragmentation as
discussed in the next section. These include a fully inclusive hadronisation
scheme according to the JETSET parton shower model~\cite{jetset} and exclusive
descriptions of the final states in the 
$b \rightarrow u \ell \bar \nu$ transition. 

\subsection{The JETSET~7.4 interface}

The generator can be used to produce individual $B$ meson decays for dedicated
studies. Timings for the generation of individual decays on different 
platforms are given in Table~1. 
BTOOL is also interfaced as an option of the LUDECY 
subroutine of JETSET to handle $b \rightarrow u \ell \bar \nu$ decays in the 
generation of $e^+ e^- \rightarrow Z^0/\gamma \rightarrow b \bar b$ events. 
The BTOOL generator can be activated by setting the MDME(IDC,2) flag in the
LUDAT3 common block of JETSET, where IDC refers to the 
$b \rightarrow u \ell \bar \nu$ decay channel.

\begin{table}[h!]
\begin{center}
\caption[]{\sl Timing of the decay generator for different platforms}

\vspace{0.2cm}

\begin{tabular}{|c|c|c|}
\hline
Platform & OS & secs./1000 decays \\
\hline \hline
HP 9000/778 & HP-UX 10.20   & 0.52 \\
%DEC 255-233 & DEC Unix V4.0 & 0.xxx \\
Pentium-133 & Linux 2.0.27  & 1.85 \\
\hline
\end{tabular}
\end{center}
\label{table:cpu}
\end{table}

\section{Results and Comparisons}

The focal interest in the simulation study of $b \rightarrow u \ell \bar \nu$ 
decays is to determine the invariant mass spectrum of the hadronic system and
its correlation with the lepton energy and then to define the uncertainties of
these distributions due to the choice of the  model and of its input 
parameters. 
Three different models have been applied for the definition of the kinematics
of the $b$ and spectator $\bar q$ inside the hadrons and two models for the 
generation of the hadronic final states. The criteria chosen for the input 
parameters and their range of variation are discussed in the following 
subsection. The results for the hadronic system mass and multiplicities 
are presented in 3.2 and 3.3.

\subsection{The Choice of Parameters}

The ACCMM model introduces two free parameters: {\it i)} the Fermi momentum 
$p_F$ and {\it ii)} the spectator quark mass $m_{sp}$. There have been 
attempts to extract the values of these parameters from fits to experimental 
observables such as the momentum
spectrum of leptons from $b \rightarrow c \ell \bar \nu$ 
decay~\cite{cleobsg,hwang}, the photon spectrum in $b \rightarrow s \gamma$
decay~\cite{alibsg} and the $J/\psi$ momentum distribution in 
$B \rightarrow J/\psi~X$~\cite{palmer}. Most of these determinations point to 
a value of $p_F \simeq$~0.5~GeV/c (Table~2).
\begin{table}[h!]
\begin{center}
\caption[]{\sl Experimental estimates of the $p_F$ value in the ACCMM model}

\vspace{0.2cm}

\begin{tabular}{|l|c|c|c|}
\hline
Channel & $p_F$ (GeV/c) & $m_{sp}$ (GeV/c$^2$) & Ref. \\
\hline \hline
$b \rightarrow c \ell \bar \nu$ & 0.27 $\pm$ 0.04 & 0.30 & 
\cite{cleobsg} \\ 
$b \rightarrow c \ell \bar \nu$ & 0.51 $^{+0.08}_{-0.07}$ & 0.0 & 
\cite{hwang} \\ 
$b \rightarrow s \gamma$ & 0.45 & 0.0 & 
\cite{alibsg} \\ 
$B \rightarrow J/\psi X$ & 0.57 & 0.15 & \cite{palmer} \\ 
\hline
\end{tabular}
\end{center}
\label{table:pfexp}
\end{table}

However it has been pointed out that the value of $p_F$ obtained in a fit to 
$b \rightarrow c$ transitions may be not appropriate in the description
of $b \rightarrow u$ decays~\cite{bigiaccmm}. At the same time it has also 
been shown that the ACCMM model is consistent with the QCD description of 
$b \rightarrow u \ell \bar \nu$ and $b \rightarrow s \gamma$ transitions and 
that the corresponding parameters can therefore be related. 
This is the direction followed in this study.

The effective $b$ quark mass $m_b$ depends on $p_F$ and $m_{sp}$ as:
\begin{eqnarray}
m^2_b = m^2_b(p_b)= m^2_B + m^2_{sp} -2 m_B \sqrt{p_b^2 + m^2_{sp}}
\end{eqnarray} 
where $p_b$ is the momentum of the heavy quark in the hadron and $m_B$ is the 
$B$ hadron mass. The value of $m_B$ can be taken as a parameter, tunable 
such that the ACCMM model corresponding to an average $b$ mass $<m_b>$ can
be compared with theory predictions obtained for a given value of $m_b$. 
Estimated values for the $b$ quark pole mass are in the range 
4.72~GeV/c$^2$~$< m_b <$~4.92~GeV/c$^2$~\cite{cleobsg}.

The value of $p_F$ is proportional to the average kinetic energy of the 
$b$ quark in the hadron since:
\begin{eqnarray}
<p_b^2> = \int^{+\infty}_{0} dp_b~p_b^2~(\phi(p_b)~p_b^2) = \frac{3}{2} p_F^2
\end{eqnarray}
where $\phi(p)$ is given by Eq.~(1).
Through the two above equations the ACCMM model parameters are related to those
of the QCD description of the heavy quark inside the hadron. In this framework
$<p_b^2>$ corresponds to the value of the expectation value $\mu_{\pi}^2$ of 
the kinetic operator. The value $p_F$ = 0.5~GeV/c corresponds to $<p_b^2>$
= 0.37~GeV$^2$. Estimates of $<p_b^2>$ have been obtained both from 
theory and fits to measured spectra in $B$ decays as discussed below. 

For the Parton Model, the fragmentation function for $b$ quarks has been 
measured at LEP. Averaging over the ALEPH, DELPHI and OPAL results, the 
fraction of the $b$ quark energy taken by the beauty hadron is
$<x_B> = 0.702 \pm 0.008$~\cite{lepew}. Also the observed shape in the 
preliminary DELPHI analysis~\cite{delphipet} was compatible with that of the 
Peterson function.
These results point to a value for the $\epsilon_b$ parameter of 
$\epsilon_b$ = 0.0040.
In the simulation of decays with the parton model, the spectator quark mass
mass $m_q$ was set to zero and the $b$ quark mass was varied in the range
4.72~GeV/c$^2$~$< m_b <$~4.92~GeV/c$^2$ as for the ACCMM model.
It is interesting to point out that, by using the central value for 
$\epsilon_b$, the parton model gave $<p_b^2> =$~0.35~GeV$^2$ for 
$m_b$~=~4.72~GeV/c$^2$, consistent with the value obtained in the ACCMM model 
for $p_F$ = 0.5~GeV/c.

The use of the QCD universal structure function $f(k_+)$ allows a consistent 
comparison with the results obtained in the framework of QCD and 
Heavy Quark expansion. The normalised moments 
$a_n = \frac{A_n}{\bar \Lambda^n}$ of $f(k_+)$, given by
\begin{eqnarray} 
a_n = \frac{1}{\bar \Lambda^n} \int d k_+ k_+^n f(k_+)
\end{eqnarray}
relate the function parameters with that of the theory. In particular the 
first two moments define the function normalisation and the third is 
proportional to the expectation value of the kinetic energy 
operator~\cite{dsu}:
\begin{eqnarray}
a_0 = 1\\
a_1 = 0\\
a_2 = \frac{3 \mu_{\pi}^2}{\bar \Lambda^2}
\end{eqnarray} 

These relationships define the values of the parameters $a$ and $c$ in Eq.~4 
as a function of the values of $\mu_{\pi}^2$ and $\bar \Lambda$. 
There have been several evaluations of $\mu_{\pi}^2$ and a selection of recent
results is given in Table~3. Results are scheme dependent and, depending on 
the method used in their derivation, they point to the values of $\mu_{\pi}^2$
of 0.4~GeV$^2$ or 0.2~GeV$^2$.

\begin{table}[h!]
\begin{center}
\caption[]{\sl Estimates of $\mu_{\pi}^2$ }

\vspace{0.2cm}

\begin{tabular}{|l|c|c|}
\hline
Method & $\mu_{\pi}^2$ (GeV$^2$) & Ref. \\
\hline \hline
QCD Sum Rules & 0.50 & \cite{ball} \\ 
Variational Method & 0.44 & \cite{hwang3} \\  
Relativ. Potential & 0.46 & \cite{deFazio} \\ 
Virial Theorem & 0.40 - 0.58 & \cite{hwang2} \\ 
Virial Theorem & 0.15 $\pm$ 0.03 & \cite{neubert3} \\ \hline
$b \rightarrow c \ell \bar \nu$ & 0.19 $\pm$ 0.10 & \cite{gremm} \\ 
$b \rightarrow c \ell \bar \nu$ & 0.14 $\pm$ 0.03 & \cite{chernyak} \\ 
$b \rightarrow s \gamma$   & 0.71$^{+1.16}_{-0.70}$ & \cite{liyu} \\ 
\hline
\end{tabular}
\end{center}
\label{table:p2}
\end{table}

The two values of $\mu_{\pi}^2 =$ 0.2~GeV$^2$ and 0.4~GeV$^2$
were chosen while $\bar \Lambda$ ranged between 0.36~GeV/c$^2$ and 
0.56~GeV/c$^2$ for 4.72~GeV/c$^2$~$< m_b <$~4.92~GeV/c$^2$ and $m_B$ = 
5.28~GeV/c$^2$. The masses of the light quarks, $u$ and spectator quark, were 
set to zero.

\subsection{The Hadronic System Mass}

The precise determination of the fraction of $b \rightarrow u \ell \bar \nu$ 
transitions yielding an hadronic system with mass $M_X$ below a given 
$M_{cut}$ value is crucial in the estimation of $|V_{ub}|$. 
\begin{figure}[ht!]
\begin{center}
\begin{tabular}{c c}
\epsfig{file=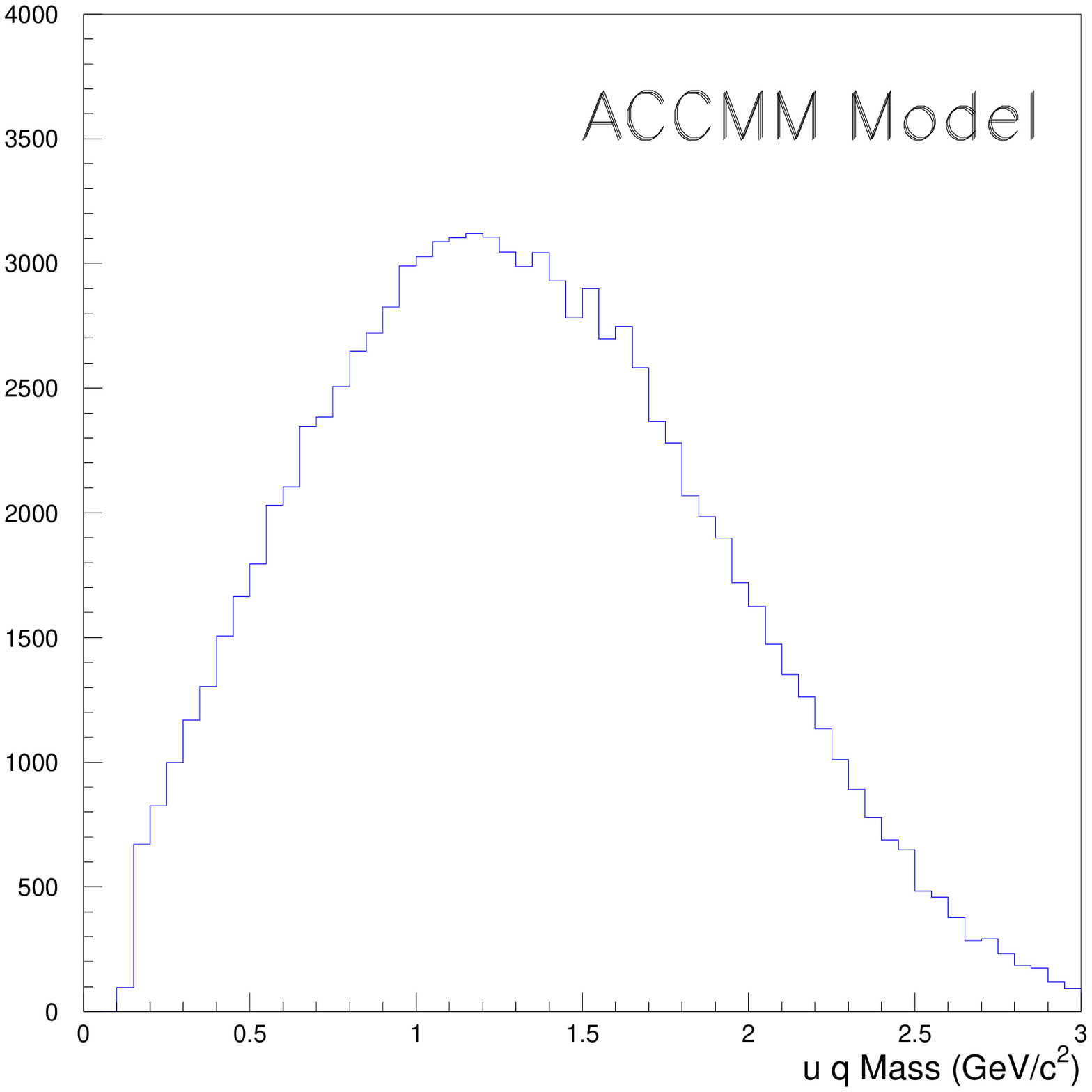,width=7.0cm,height=5.0cm,clip=} &
\epsfig{file=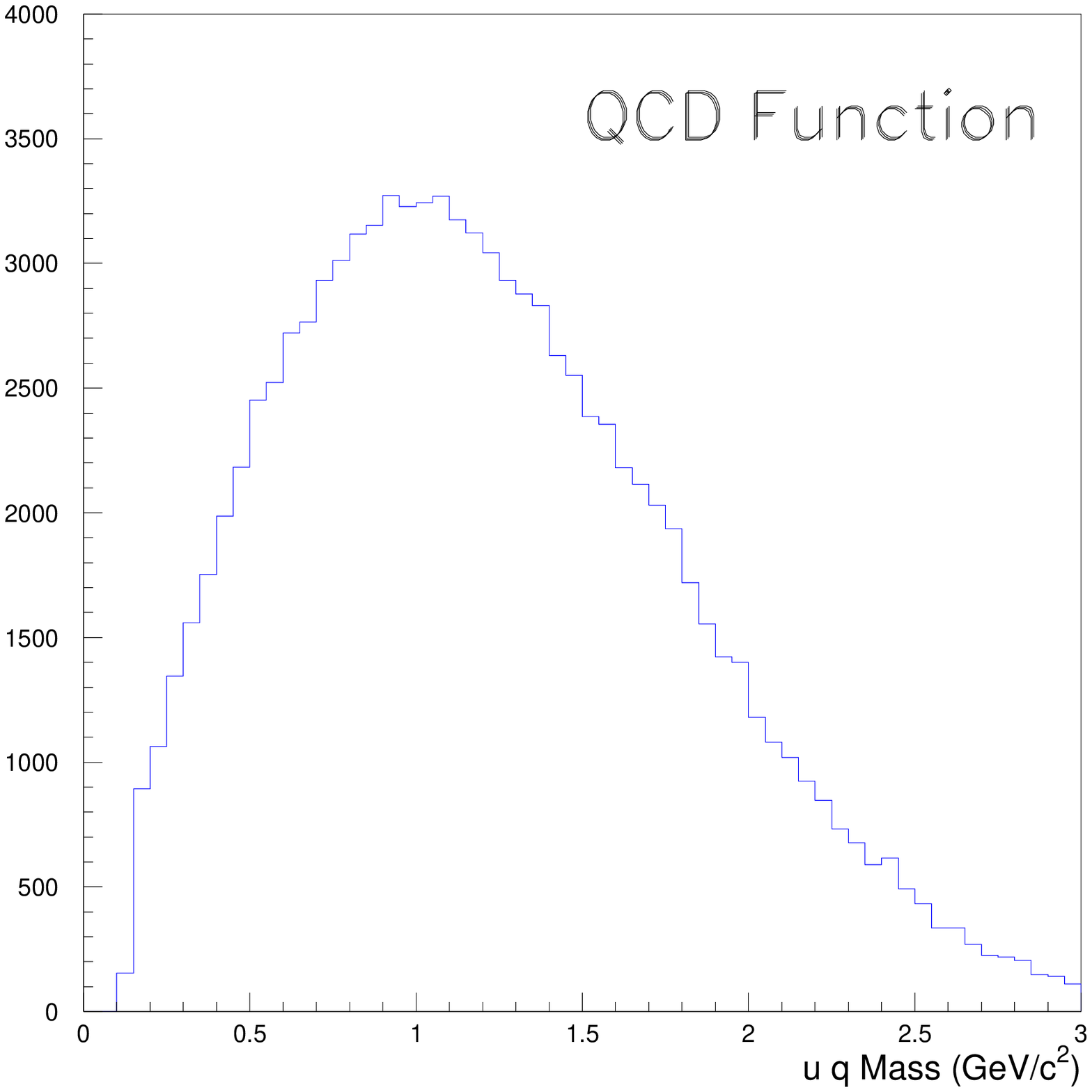,width=7.0cm,height=5.0cm,clip=} \\
\end{tabular}
\end{center}
\caption[]{\sl The invariant mass distribution of the $u \bar q$ system 
obtained using the ACCMM model (left) and the QCD universal function (right)
corresponding to $m_b$ = 4.72~GeV/c$^2$ and $<p_b^2>$ = 0.4 GeV$^2$.}
\label{fig:hadmp}
\end{figure}
As discussed above, at the quark level the hadronic mass corresponds to the 
mass of the $u \bar q$ system. This depends mainly on the masses and motion of
the heavy and spectator quarks in the $B$ hadron. On the contrary the 
experimentally measureable hadronic mass is strongly affected by the resonant 
decomposition of the hadronic final states. 
In the simulation the hadronic system is analyzed in two steps.
 
Firstly, the $u \bar q$ pair is analyzed. This gives the hadronic mass 
distribution before resonant states are taken into account (see Figure~2) and 
can be compared with that computed using QCD and heavy quark 
expansion~\cite{bdu}. 

Results are expressed in terms
of the fraction $F_u(M_{cut})$ of the $b \rightarrow u \ell \bar \nu$ 
transitions 
resulting in a mass of the $u \bar q$ pair below a given cut value. 
The kinematics of the $b$ and spectator quark have been defined
using the ACCMM model, the QCD universal structure function and the Parton 
Model. In order to compare the results, parameters have been chosen such that
they correspond to $m_b$~=~4.80~$\pm$~0.10~GeV/c$^2$ and 
0.2~GeV$^2$~$<~<p_b^2>~<$~0.4~GeV$^2$ as discussed above.

\begin{table}[hb!]
\begin{center}
\caption[]{\sl Fraction $F_u(M_{cut})$ of $b \rightarrow u \ell \bar \nu$ 
decays with $M_X < M_{cut}$ for $<p_b^2>$ = 0.4~GeV/c$^2$.}

\vspace{0.2cm}

%\begin{tabular}{|c|c|c|c|c|c|c|}
%\hline
%$m_b$ & $<p_b^2>$ & $M_{cut}$ & $F_u(M_{cut})$ & $F_u(M_{cut})$ & 
%$F_u(M_{cut})$ & $F_u(M_{cut})$ \\
%(GeV/c$^2$) & (GeV$^2$) & (GeV/c$^2$) & ACCMM & QCD Funct. & Parton &
%Ref.~\cite{bdu}\\
%\hline \hline
%4.92 & 0.40 &         1.25 & 0.48 & 0.50 & 0.40 & 0.45 \\
%\mbox{ } & \mbox{ } & 1.50 & 0.65 & 0.60 & 0.62 & 0.60 \\
%\mbox{ } & \mbox{ } & 1.75 & 0.78 & 0.80 & 0.84 & 0.78 \\
%\hline
%4.82 & 0.40 &         1.25 & 0.50 & 0.57 & 0.45 & 0.55 \\
%\mbox{ } & \mbox{ } & 1.50 & 0.65 & 0.72 & 0.74 & 0.72 \\
%\mbox{ } & \mbox{ } & 1.75 & 0.76 & 0.83 & 0.89 & 0.85 \\
%\hline
%\end{tabular} 

\begin{tabular}{|c|c|c|c|c|c|}
\hline
$m_b$ & $M_{cut}$ & $F_u(M_{cut})$ & $F_u(M_{cut})$ & 
$F_u(M_{cut})$ & $F_u(M_{cut})$ \\
(GeV/c$^2$) & (GeV/c$^2$) & ACCMM & QCD Funct. & Parton &
Ref.~\cite{bdu}\\
\hline \hline
4.92     & 1.25 & 0.48 & 0.50 & 0.40 & 0.45 \\
\mbox{ } & 1.50 & 0.65 & 0.60 & 0.62 & 0.60 \\
\mbox{ } & 1.75 & 0.76 & 0.80 & 0.84 & 0.78 \\
\hline
4.82     & 1.25 & 0.50 & 0.57 & 0.45 & 0.55 \\
\mbox{ } & 1.50 & 0.65 & 0.72 & 0.74 & 0.72 \\
\mbox{ } & 1.75 & 0.78 & 0.83 & 0.89 & 0.85 \\
\hline
4.72     & 1.25 & 0.54 & 0.65 & 0.47 & 0.68 \\
\mbox{ } & 1.50 & 0.68 & 0.75 & 0.71 & 0.81 \\
\mbox{ } & 1.75 & 0.80 & 0.85 & 0.87 & 0.89 \\
\hline
\end{tabular} 

\end{center} 
\label{table:hadmass}
\end{table}

Table~4 summarizes the results for different choices of the input parameters.
The first observation is that the universal function implemented in 
BTOOL reproduces the prediction from QCD and heavy quark expansion 
from~\cite{bdu}. Further it has been found that also the ACCMM model 
reproduces these results to better than 15$\%$, for equivalent values of 
$m_b$ and $<p_b^2>$. This study confirms that the sensitivity to the value of 
the $b$ quark mass is significant in the region of low hadronic invariant 
masses. In this region also the model dependence is more pronounced.
In order to estimate the overall uncertainty in the estimate of the fraction 
of $b \rightarrow u \ell \bar \nu$ decays with hadronic mass below a given cut
value $M_{cut}$, the different sources of systematics have been combined.
The chosen range of variation of the parameters is 
$\pm \sigma(m_b)$ = $\pm$ 0.10~GeV/c$^2$ and $\pm \sigma (<p_b^2>)$ = 
$\pm$ 0.1~GeV$^2$. The corresponding realtive systematic errors are 
summarized in Table~5.
\begin{table}[h!]
\begin{center}
\caption[]{\sl Estimate of the relative systematic uncertainties on 
$F_u(M_{cut})$ }

\vspace{0.2cm}

\begin{tabular}{|l|c|c|c|}
\hline
Source & $M_{cut}$ = 1.25 &  $M_{cut}$ = 1.50 &  $M_{cut}$ = 1.75 \\
\hline \hline
$m_b$     & 0.13 & 0.10 & 0.05 \\ 
$<p_b^2>$ & 0.08 & 0.04 & 0.02 \\  
Model     & 0.08 & 0.04 & 0.04 \\
\hline 
{\bf Total } & 0.17 & 0.12 & 0.07 \\
\hline
\end{tabular}
\end{center}
\label{table:syst}
\end{table}

Secondly, the hadronic final states corresponding to a given mass and energy 
of the $u \bar q$ pair can be predicted by a variety of methods ranging
from the fully inclusive quark fragmentation approach to exclusive models. 

\begin{figure}[hb!]
\begin{center}
\epsfig{file=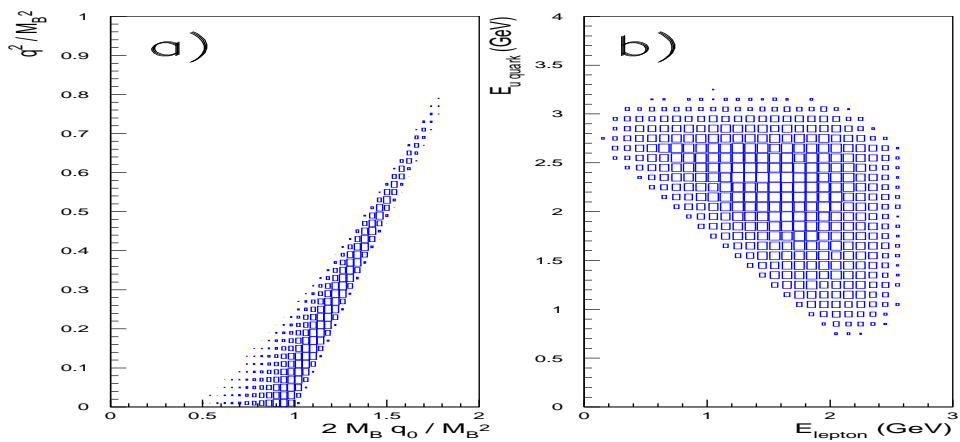,width=14.0cm,height=6.5cm,clip=}
\end{center}
\caption[]{\sl The distribution of the $u$ quark energy versus the lepton 
energy (a) and the physical region in the $q^2$ versus $2 m_B q_0$ plane (b)
for inclusive $b \rightarrow u \ell \bar \nu$ decay.}
\label{fig:dalitz}
\end{figure}
At large enough recoil $u$ quark energies, the $u \bar q$ system moves away 
fast and this picture is similar to that of the evolution of a jet initiated
by a light quark $q$ in $e^+e^- \rightarrow q \bar q$ annihilation. This is 
simulated by first arranging the $u \bar q$ system in a string configuration 
and then making it fragment according to the parton shower model. 
Exclusive models compute the decay amplitudes from the heavy-to-light form
factors and the quark hadronic wave functions. The so-called ISGW2 
model~\cite{isgw2} approximates the inclusive 
$b \rightarrow u \ell \bar \nu$ decay
width by the sum over resonant final states, taking into account leading 
corrections to the heavy quark symmetry limit.

The range of applicability of the inclusive and exclusive models is restricted
to particular regions of the accessible kinematic configurations. Figure~3~a) 
shows the physical region
in the $q^2$ versus $2~m_B~q_0$ plane where $q^2$ is the effective mass of the
virtual $W$ and $q_0$ its energy in the $B$ rest frame. In this plot states of
equal hadronic invariant mass $M_X$ correspond to lines 
$M_X^2 = m_B^2 - 2 m_B q_0 + q^2$. 
Systems of large invariant mass correspond to low $q^2$ and $q_0$ values, i.e.
large $u$ recoil as shown in Figure~3~b). In these cases the 
$u$~quark energy 
is typically large enough compared with that of the spectator quark that 
the analogy with jet fragmentation is justifiable. Conversely at low $u$ 
recoil energy, i.e. close to the upper kinematical limit in 
Figure~3~a), the relative momentum of the $u \bar q$ pair is 
small and they are therefore likely to form a bound state. 
\begin{figure}[hb!]
\begin{center}
\epsfig{file=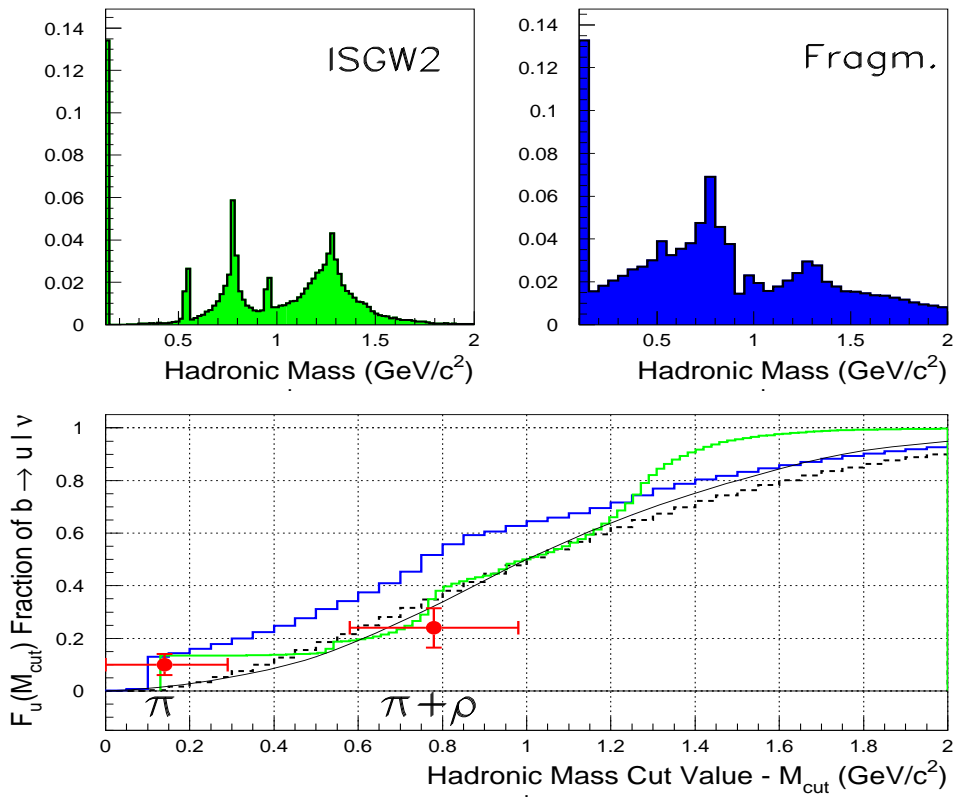,width=14.0cm,height=12.0cm,clip=}
\end{center}
\caption[]{\sl Comparison of the invariant mass of the hadronic final state
from the ISGW2 model (upper left) and from parton shower fragmentation 
(upper right). In the lower plot the corresponding fractions $F_u(M_{cut})$ of
the $b \rightarrow u \ell \bar \nu$ transitions resulting in a mass below a 
given cut value $M_{cut}$ are shown as a function of $M_{cut}$. The 
corresponding predictions for the invariant mass of the $u \bar q$ pair are 
shown with the continuous line (\cite{bdu}) and the dash histogram (BTOOL). 
The CLEO results for the branching ratios for the $\pi \ell \bar \nu$ and 
$\rho \ell \bar \nu$ channels are also shown for comparison.}
\label{fig:isgw}
\end{figure}

A satisfactory description of the inclusive $b \rightarrow u \ell \bar \nu $ 
decay must take these characteristics into account. Hybrid models have been 
proposed for this purpose~\cite{hybrid}. The main feature of an hybrid model 
is to define a kinematical region in which the exclusive model is valid and 
its complement that can be treated by an inclusive fragmentation model. The 
two regions must be chosen in order to have well behaved matching conditions 
for a set of relevant kinematical variables. 
Further constraints can be derived from the branching ratios for 
$B \rightarrow \pi \ell \bar \nu$ and $B \rightarrow \rho \ell \bar \nu$ 
decays measured by CLEO~\cite{cleo2,cleo3}.
Since the aim of this study is the definition of the systematic uncertainties 
in the description of the hadronic mass spectrum, results have been derived 
for the two extreme cases of fully inclusive and exclusive models. In the 
inclusive model,
the probabilities for generating light vector and axial resonances have been 
tuned in JETSET in order agree with the measured rates in $Z^0$ decays. 
For the exclusive model the ISGW2 model has been used. 
The results are presented in Figure~4 in terms of the fraction of 
$b \rightarrow u \ell \bar \nu$ decays with the hadronic final state below a 
given mass value. The comparison of the predictions from the two models shows 
significant differences in the predicted mass spectra due to the relative 
importance of resonant and non-resonant final states. For the region of
$M_{cut} >$ 1.5~GeV/$c^2$, the relative difference of the two models 
correponds to 10 - 15\% showing that the hadronic system fragmentation 
introduces an uncertainty of the same order as that from the $b$-quark mass and
the heavy hadron kinematics. 

\subsection{The Hadronic System Multiplicity}

An additional source of uncertainty in the modelling of the decay arises from
the multiplicity of the hadronic system. This is of special relevance since
the efficiency for resonstructing the decay depends on this multiplicity. In 
order to study this uncertainty, the different prescriptions for describing 
the final states discussed above have been analyzed in terms
of the resulting decay multiplicity.

\begin{figure}[hb!]
\begin{center}
\epsfig{file=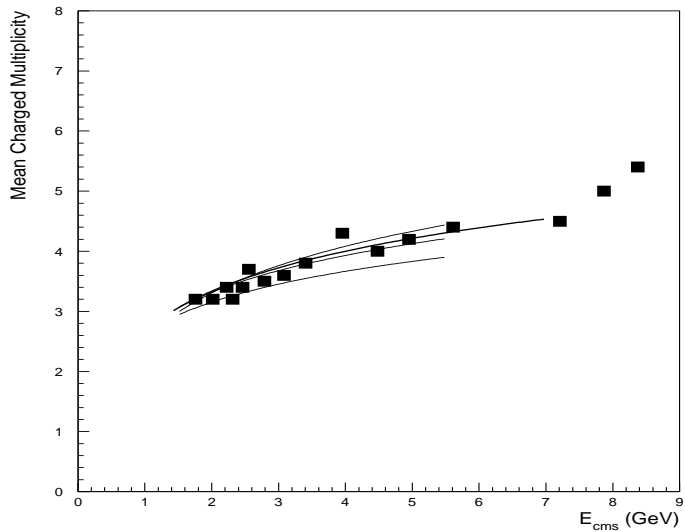,width=10.0cm,height=8.0cm,clip=}
\end{center}
\caption[]{\sl Comparison of the generator predictions with 
experimental data on charged multiplicity in low-energy $e^+e^-$ collisions.
The points represent the experimental data, the long thick line the best fit 
from Ref.~\cite{Hollebeek} and the shorter thinner lines the fits to the 
generator results for values of $p_F$ ranging from 0.0~GeV/c to 0.35~GeV/c.}
\label{fig:mult}
\end{figure}
\newpage

The charged multiplicity of the hadronic system from the 
$b \rightarrow u \ell \bar \nu$ decay can be 
compared with one half of that of a $q \bar q$ event at $\sqrt{s} = 2 E_{had}$.
The data on the event charged multiplicity from {\sc Adone} and {\sc Mark~II} 
at 2~GeV~$< \sqrt{s} <$~8~GeV can be described with a function 
$<n_{ch}> = a + b~ln~s$ with $a=2.67 \pm 0.04$ and 
$b=0.48 \pm 0.02$~\cite{Hollebeek} shown by the long thick line in Figure~5. 
These data have also been compared with the 
results of the simulation where a $u$ quark of energy $\sqrt{s}/2$ is paired 
with the spectator quark. The quark is either given Fermi motion according to
the value of $p_F$ but no transverse momentum, or kept at rest (i.e. $p_F=0$).
The results are shown by the shorter thinner lines in Figure~5. The parton 
shower model reproduces reasonably well both the multiplicity and its scaling 
with the quark energy. The best agreement with the data is obtained by 
imposing $p_F$ = 0.2~GeV/c which gives a fit with $a$=2.71 and $b$=0.44.

Multiplicities in semileptonic $B$ decays  have also been predicted using the 
quark-gluon string model (QGSM) which also reproduces fairly precisely the 
same data on the charged event multiplicity in low energy 
$e^+e^-$~collisions~\cite{dobrov}. The multiplicities obtained by the decay 
generator using the hybrid model are compared with the predictions from QGSM 
and IGSW2 in Table~6. The average charged multiplicity in the decay
$<n_{ch}>$ agrees for the three models within $\pm$~0.16. This multiplicity 
is also quite close to that measured for $D$ meson decays~\cite{markiii}, 
showing that the reconstruction of the hadronic system may be performed with 
comparable efficiency in semileptonic $b \rightarrow u$ and 
$b \rightarrow c$ decays.

\begin{table}[ht!]
\begin{center}
\caption[]{\sl Decay multiplicities from the event generator compared with 
QGSM and IGSW2 models} 

\vspace{0.2cm}

\begin{tabular}{|l|c|c|c|}
\hline
Final State & Simulated & QGSM~\cite{dobrov} & ISGW2~\cite{isgw2} \\
\hline \hline
$0~\pi^{\pm}~~~1~\pi^0$ & .050 & .020 & .044 \\
$0~\pi^{\pm}~~~2~\pi^0$ & .016 & .002 & .010 \\
$0~\pi^{\pm}~~~3~\pi^0$ & .011 & .009 & .012 \\
$0~\pi^{\pm}~~~4~\pi^0$ & .005 & .004 & .004 \\
\hline
$1~\pi^{\pm}~~~0~\pi^0$ & .101 & .060 & .090 \\
$1~\pi^{\pm}~~~1~\pi^0$ & .124 & .140 & .135 \\
$1~\pi^{\pm}~~~2~\pi^0$ & .037 & .064 & .016 \\
$1~\pi^{\pm}~~~3~\pi^0$ & .008 & .025 & .008 \\
$1~\pi^{\pm}~~~4~\pi^0$ & .003 & .003 &   -  \\
\hline
$2~\pi^{\pm}~~~0~\pi^0$ & .117 & .080 & .090 \\
$2~\pi^{\pm}~~~1~\pi^0$ & .176 & .175 & .167 \\
$2~\pi^{\pm}~~~2~\pi^0$ & .038 & .095 & .048 \\
$2~\pi^{\pm}~~~3~\pi^0$ & .004 & .015 & .005 \\
\hline
$3~\pi^{\pm}~~~0~\pi^0$ & .053 & .070 & .054 \\
$3~\pi^{\pm}~~~1~\pi^0$ & .062 & .098 & .061 \\
$3~\pi^{\pm}~~~2~\pi^0$ & .021 & .036 &   -  \\
\hline
$4~\pi^{\pm}~~~0~\pi^0$ & .012 & .050 & .006 \\
$4~\pi^{\pm}~~~1~\pi^0$ & .017 & .020 & .003 \\
\hline
$<n_{ch}>$ & 1.91 & 2.06 & 1.70 \\
\hline
\end{tabular}
\end{center}
\label{table:mult}
\end{table}

\section{The Lepton Spectrum}

As already mentioned, the lepton spectrum is 
sensitive to the mass of the quark produced in the semileptonic $b$ decay.
While the lepton yield in the region of lepton energies above the kinematical 
limit $\simeq \frac{M_B^2 -M_D^2}{2 M_B}$ for $b \rightarrow c \ell \bar \nu$
transitions is subject to significant model dependences, a combined study 
of the mass of the hadronic system $M_X$ and the energy of the lepton in
the $B$ rest frame $E^*_{\ell}$ may allow an extraction of the $|V_{ub}|$ with 
good sensitivity and improved control of the systematics. 
\begin{figure}[hb!]
\begin{center}
\begin{tabular}{c c}
\epsfig{file=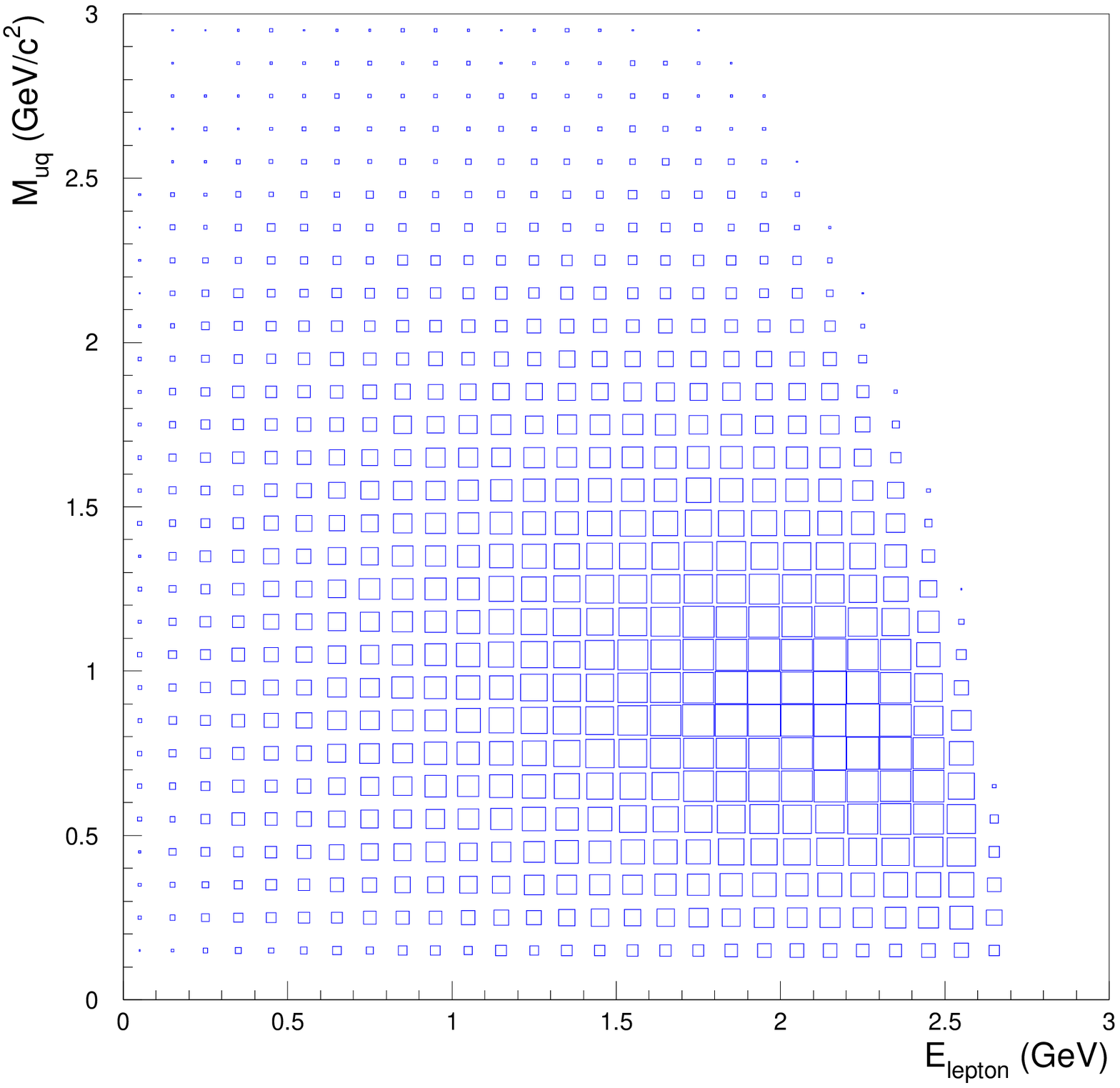,width=7.0cm,height=6.5cm,clip=} &
\epsfig{file=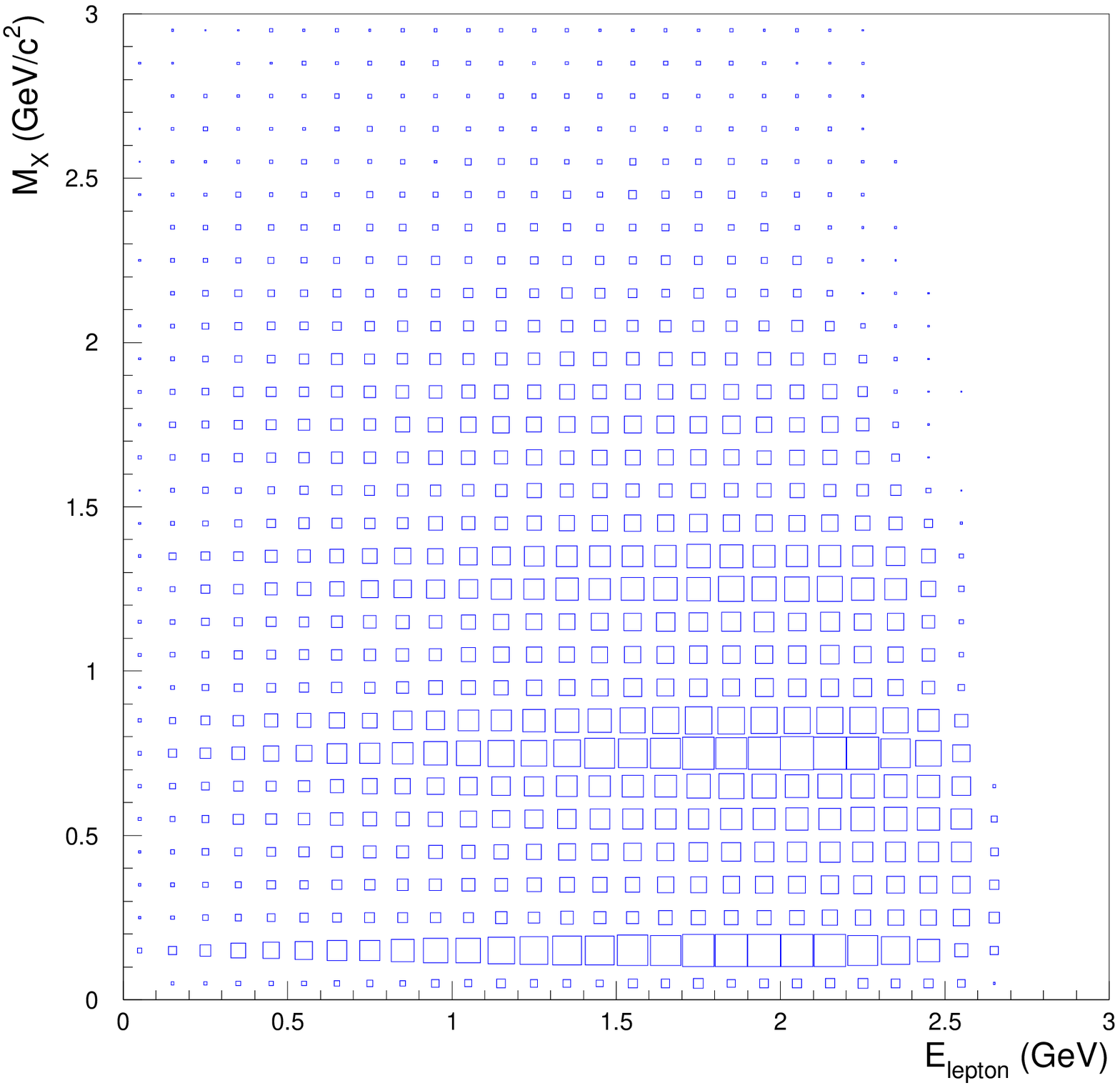,width=7.0cm,height=6.5cm,clip=}\\
\end{tabular}
\end{center}
\caption[]{\sl Invariant mass of the hadronic system vs. lepton energy in 
$b \rightarrow u \ell \bar \nu$ decays for the $u \bar q$ system (left) and 
the hadronic final state (right).}
\label{fig:lept1}
\end{figure}
Figure~6 shows the 
correlation between the values of $E^*_{\ell}$ and $M_X$ before and after the 
hadronisation of the $u \bar q$ system. By selecting decays with low hadronic
invariant mass, the lepton spectrum is depleted in its lower end without 
significantly affecting the region of lepton energies above 1.5~GeV/c that
are relevant for separating $b \rightarrow u \ell \bar \nu$ from 
$b \rightarrow c \ell \bar \nu$ decays. 

\section{Conclusion}

A generator for inclusive $b \rightarrow u \ell \bar \nu$ decays has been 
developed and used for studying the invariant mass and resonance decomposition
of the hadronic system produced in the decay.
These studies are of special relevance for the extraction of $|V_{ub}|$ from 
semileptonic $B$ decays at LEP and at the $B$ factories. The low invariant 
mass of the hadronic system emitted in these decays can be used to separate 
them from the CKM favoured $b \rightarrow c \ell \bar \nu$ transitions. 
Different models for defining the 
kinematics of the heavy quark inside the hadron and the $u \bar q$ system 
hadronisation have been compared. Systematic uncertainties in the fraction of
decays giving an hadronic system with mass below a given cut arise from the 
value of the $b$ quark pole mass, the momentum distribution of the heavy and
spectator quark inside the hadron and modelling of the $u \bar q$ 
hadronisation.  This analysis confirmed the observation that model dependencies
and these systematic uncertainties from the $b$-quark mass and the heavy quark
kinematics can be kept at the 10\% level, or below,
if the study of $b \rightarrow u \ell \bar \nu$ is performed including decays 
with hadronic final state masses up to $\simeq$ 1.6~GeV/c$^2$ or above.
A comparable uncertainty arises from the hadronisation model when comparing an
inclusive to a fully exclusive model.
The combined analysis of the hadronic mass and lepton spectrum spectrum may
provide an optimal separation of $b \rightarrow u \ell \bar \nu$ from 
$b \rightarrow c \ell \bar \nu$ decays.
 
\vspace{1.0cm}

\noindent
{\Large \bf Acknowledgements}

\vspace{0.5cm}

\noindent
I would like to thank M.~Neubert and N. Uraltsev for extensive discussions.
I am also grateful to I.~Bigi and C.S.~Kim for pointing out the relationships 
between ACCMM and QCD parameters, to D.~Lange and A.~Ryd for providing me 
with results from a Monte Carlo implementation of the ISGW2 model, 
to T.~Sj\"ostrand for
his advices on interfacing this generator with JETSET and to W. Venus for his
comments to the manuscript.

%=========================================================================%

\end{document}